\documentclass[review,12pt]{elsarticle}
\usepackage{graphicx}
\usepackage{amsmath }
\usepackage{lineno}
\usepackage{upgreek}
\usepackage{subcaption}
\usepackage{natbib}
\biboptions{sort&compress}

\journal{XXXXXX}

\begin{document}
	
	\begin{frontmatter}
		
		\title{Temporal dynamics of all-optical switching in hybrid VO$_{2}$/Si waveguides}

		\author[1]{Jorge Parra}
		%\ use * to mark the author as the corresponding author
		\author[1]{Todora Ivanova}
		\author[2,3]{Mariela Menghini} 
		\author[2]{Pía Homm}
		\author[2]{Jean-Pierre Locquet}
		\author[1]{Pablo Sanchis$^{*,}$}
		\address[1]{\protect\raggedright 
			Nanophotonics Technology Center, Universitat Politènica de València, Camino de Vera s/n, 46022 Valencia, Spain, $^{*}$e-mail: pabsanki@ntc.upv.es}
		\address[2]{\protect\raggedright 
			Department of Physics and Astronomy, KU Leuven, Celestijnenlaan 200D, 3001 Leuven, Belgium}
		\address[3]{\protect\raggedright 
			IMDEA Nanociencia, Calle Faraday 9, E28049, Madrid, Spain}
		
		\begin{abstract}
			Vanadium dioxide (VO$_{2}$) is one of the most promising materials for developing hybrid photonic integrated devices (PICs). However, despite switching times as low as a few femtoseconds have been reported, the all-optical temporal dynamics of VO$_{2}$ embedded in a waveguide using an in-plane optical signal remain still hidden. Here, we experimentally investigate this behavior in hybrid VO$_{2}$/Si waveguides by using pump-probe measurements at telecom wavelengths. Our results show switching times in the micro and nanosecond range, suggesting that the phase transition is triggered thermally from the light absorbed by the VO$_{2}$ and the temporal response is governed by thermal conductive dynamics. By properly engineering the optical pulse, we prospect switching times of nanoseconds with an energy consumption of a few nanojoules. Our results unveil a new temporal dynamic that would be useful for developing future all-optical VO$_{2}$ photonic integrated devices.
		\end{abstract}
		
		
	\end{frontmatter}
	
%\keywords{optical switching, photonic integration, vanadium dioxide, silicon photonics.}

\section{Introduction} 

Phase change materials (PCMs) present unique optical properties that can be tuned by external stimuli \cite{Morin1959}. Among them, vanadium dioxide (VO$_{2}$) is a complementary metal-oxide-semiconductor (CMOS)-compatible material that exhibits a reversible and hysteretic insulating-to-metal transition (IMT)  \cite{Liu2018}.  Such IMT can be induced with different external stimuli such as temperature \cite{VanBilzen2015}, electric field \cite{Chae2005,Ko2008} or laser-induced \cite{Becker1996}.

Notably, the IMT is accompanied with a significant change of the optical properties at telecom wavelengths. The refractive index, $n+j\kappa$, of VO$_{2}$ goes from $3.21+j0.17$ in the insulating state to $2.15+j2.79$ in the metallic one at 1550 nm \cite{Briggs2010}. This large difference in both real and imaginary parts has been exploited in the silicon (Si) photonics platform for developing ultra-compact hybrid VO$_{2}$/Si devices \cite{Miller2018,Miller2017,Shibuya2019a,Sanchez2018,Markov2015,Joushaghani2015}. Different switching timescales can be obtained depending on the stimuli used on the device. A few microseconds are obtained using microheaters \cite{Sanchez2018} and it can be decreased to nanoseconds by electrical triggering \cite{Markov2015,Joushaghani2015}.

However, the optical approach turns out as the fastest way to induce the IMT of VO$_{2}$ as switching times between femtoseconds and several picoseconds have been reported \cite{Becker1996,Cavalleri2001,Cavalleri2004,Baum2007,Bionta2018,Haglund2019}. The origin of such ultra-fast switching times is not yet fully clear due to the complex interplay between the lattice and electronic degrees of freedom involved in the VO$_{2}$ phase change \cite{Otto2019}. Previous works have been focused on the mechanism behind the photo-induced phase transition of VO$_{2}$ using ultra-fast pump-probe studies. However, the  temporal dynamics are retrieved by optically exciting the VO$_{2}$ out of plane. Recently, Wong \emph{et al.} have demonstrated for the first time in-plane all-optical switching using a VO$_{2}$/SiN waveguide \cite{Wong2019}. However, they did not measure the temporal response of the device, thus leaving still unaddressed such useful topic for developing new all-optical hybrid VO$_{2}$-waveguide devices.

Here, we experimentally report temporal measurements of in-plane and all-optical switching in hybrid VO$_{2}$/Si waveguides. To this end, we retrieved the temporal response of these waveguides on transmission upon optical pulses using a pump-probe technique in the telecommunications wavelength region.

\section{Device and VO$_{2}$ characterization}
\begin{figure}[!ht]
	\centering
	\begin{subfigure}[c]{0.49\linewidth}
		\centering
		\includegraphics[width=\linewidth]{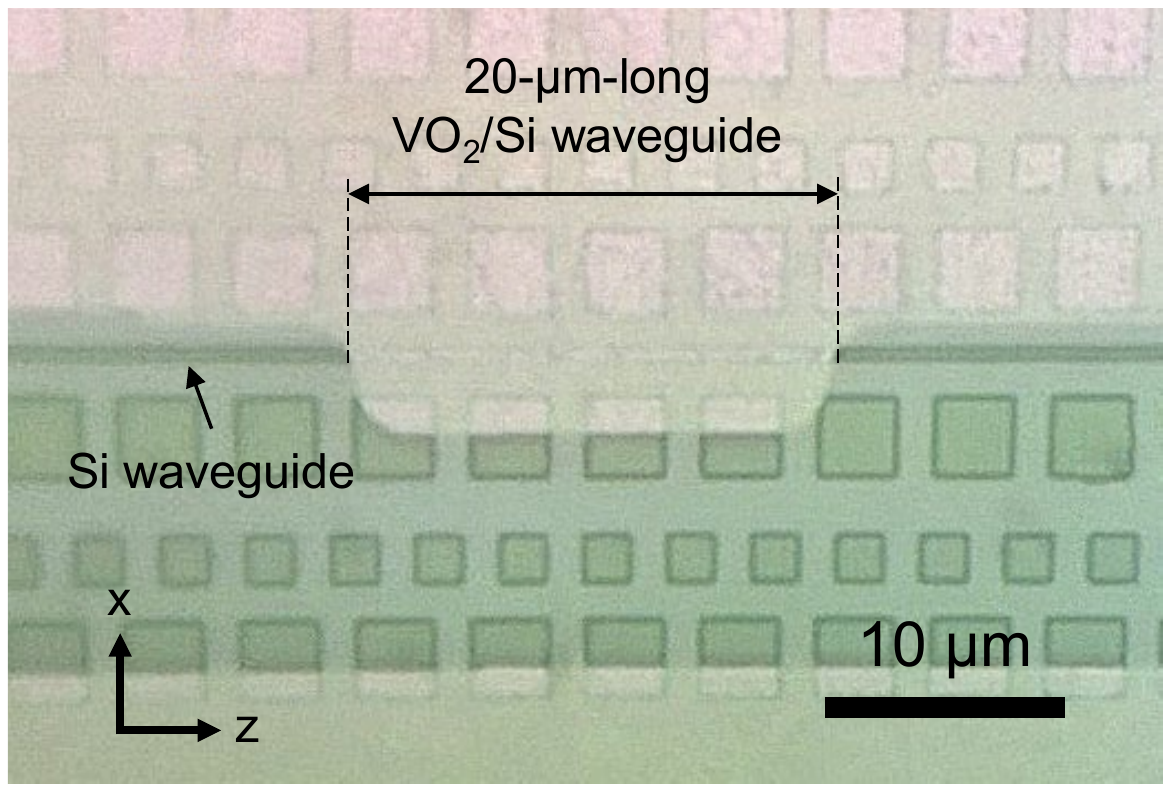}
		\caption{\centering}
		\label{fig:VO2-WG-Optical}
	\end{subfigure}	
	\hfill
	\begin{subfigure}[c]{0.49\linewidth}
		\centering
		\includegraphics[width=\linewidth]{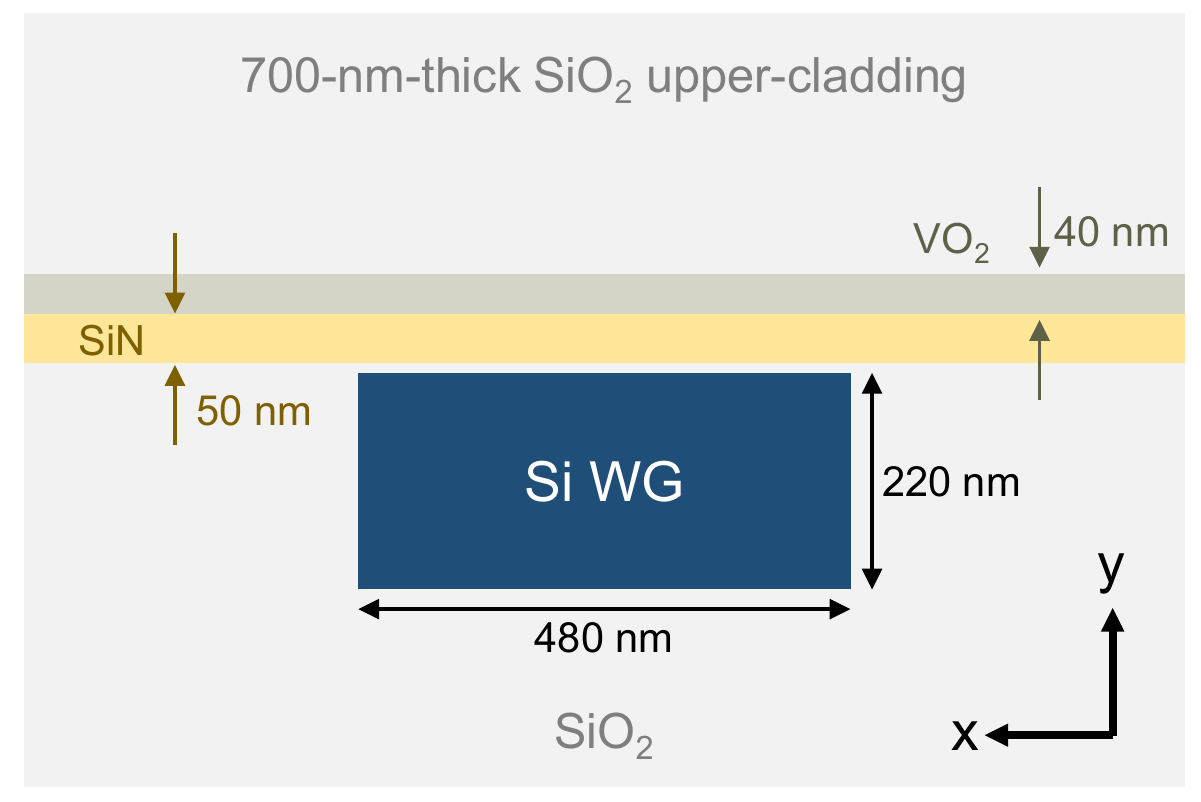}
		\caption{\centering}
		\label{fig:Schematic}
	\end{subfigure}	
	\caption{(a) Optical image of the fabricated hybrid 20-$\upmu$m-long VO$_{2}$/Si waveguide before depositing the SiO$_{2}$ upper-cladding. (b) Cross-section scheme of the hybrid waveguide.}
\end{figure}

The fabricated hybrid VO$_{2}$/Si waveguide used in this work is shown in Fig. \ref{fig:VO2-WG-Optical}. This comprises  a Si waveguide with a 20-$\upmu$m-long VO$_{2}$ patch on top. A scheme of the cross-section is depicted in Fig. \ref{fig:Schematic}. The Si waveguide is 480 nm $\times$ 220 nm to work in the single-mode region. A 50-nm-thick SiN layer is used for planarization of the silicon surface. Between the silicon waveguide and the SiN layer there is a 10-nm-thick of oxide. A 40-nm-thick VO$_{2}$ layer is placed atop of the SiN layer. The VO$_{2}$ layer was formed by growing a VO$_{x}$ layer by molecular beam epitaxy (MBE) followed by an ex-situ post-annealing at 400 ºC in forming gas \cite{VanBilzen2015}. Finally, a 700-nm-thick SiO$_{2}$ upper-cladding was deposited by plasma-enhanced chemical vapor deposition (PECVD) to avoid oxidation of the VO$_{2}$ due to environmental variations.

\begin{figure}[!ht]
	\centering
	\begin{subfigure}[c]{0.49\linewidth}
		\centering
		\includegraphics[width=\linewidth]{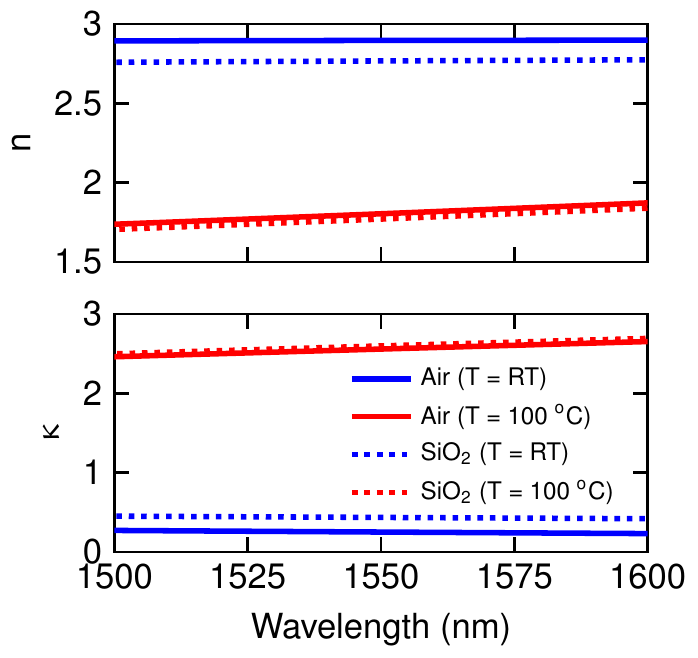}
		\caption{\centering}
		\label{fig:Ellipsometry}
	\end{subfigure}	
	\hfill
	\begin{subfigure}[c]{0.49\linewidth}
		\centering
		\includegraphics[width=\linewidth]{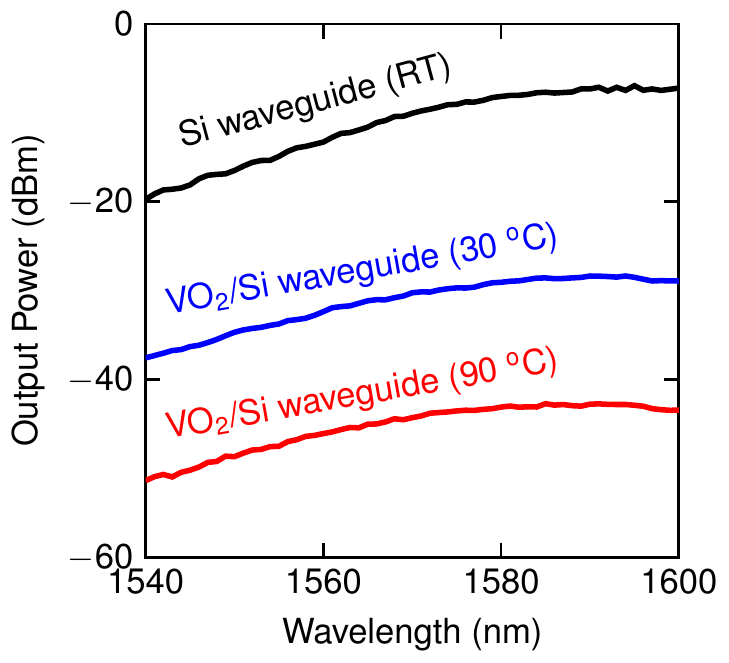}
		\caption{\centering}
		\label{fig:Spectra}
	\end{subfigure}	
	\\
	\begin{subfigure}[c]{0.9\linewidth}
		\centering
		\includegraphics[width=\linewidth]{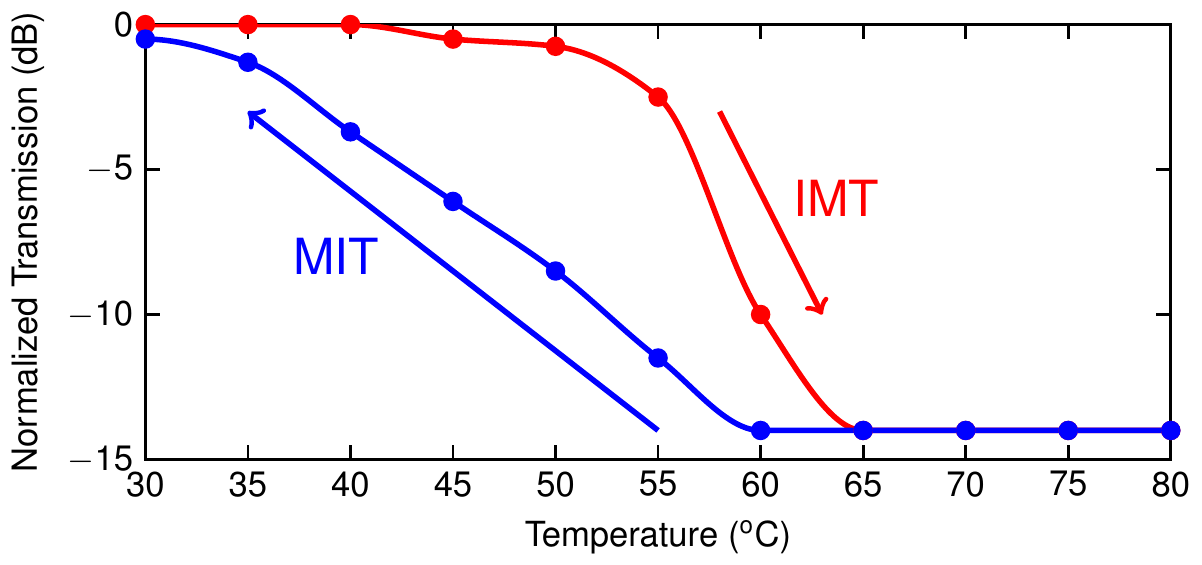}
		\caption{\centering}
		\label{fig:Hysteresis}
	\end{subfigure}	
	\caption{(a) Refractive indices of the VO$_{2}$ layer in the insulating (T = RT) and metallic (T = 100 ºC) states obtained by spectroscopic ellipsometry with an air and SiO$_{2}$ cladding. (b) Spectra of a reference silicon waveguide and the hybrid waveguide in the insulating (T = 30 ºC) and metallic (T = 90 ºC) states. Laser output power was 0 dBm. (c) Normalized transmission of the hybrid waveguide as a function of the temperature for a heating-cooling cycle. Results are given at 1550 nm.}
\end{figure}

Spectroscopic ellipsometry measurements were carried out on a dummy Si sample with the same 40-nm-thick VO$_{2}$ at room temperature (RT) and 100 ºC by using a hot plate to determine the VO$_{2}$ refractive index both in the insulating and metallic state. These measurements were performed before and after depositing the SiO$_{2}$ cladding to check for possible variations in the refractive index of the VO$_{2}$ due to this processing step.. Similar values were obtained as shown in Fig. \ref{fig:Ellipsometry}, so the quality of the VO$_{2}$ was not compromised. Its refractive index varied from $2.74 + j0.5$ to $1.78 + j2.58$ at 1550 nm between the insulating and metallic state, respectively, in good agreement with the literature \cite{Briggs2010}. 

Based on these results, the effective refractive index, $n_{\text{eff}} + j\kappa_{\text{eff}}$, was calculated at 1550 nm and transverse electric (TE) polarization for both VO$_{2}$ states by finite element method (FEM). These were $2.47 + j0.02$ and $2.41 + j0.04$ for the VO$_{2}$ in the insulating and metallic state, respectively. From the value of $\kappa_{\text{eff}}$, the optical loss induced by the 20-$\upmu$m-long VO$_{2}$ patch was obtained, leading to 14.08 dB (28.16 dB) in the insulating (metallic) state.

The spectrum of the hybrid waveguide was measured with the VO$_{2}$ in the insulating and metallic state (Fig. \ref{fig:Spectra}). To this end, the temperature of the chip was set at 30 ºC and 90 ºC, respectively, by using a Peltier device. Light was coupled to/from the chip by means of TE grating couplers. A reference waveguide without VO$_{2}$ was also measured to estimate grating and VO$_{2}$ losses. Experimental optical loss of VO$_{2}$ in the insulating and metallic state was 18.2 dB and 32.2 dB, respectively, which delivered an extinction ratio of 14 dB, in fair agreement with simulations.

The typical hysteretic response of VO$_{2}$ on its optical loss was obtained upon a heating-cooling cycle (Fig. \ref{fig:Hysteresis}). On one hand, the IMT occurred around 58 ºC in a window between 55 ºC and 63 ºC. On the other hand, the metal-to-insulating transition (MIT) started at 60 ºC and finished around 30 ºC. This difference in the steepness between the IMT and MIT, being the first more abrupt, has also been reported in previous works \cite{Briggs2010} and it is attributed to different crystal grain sizes during the structural phase transition \cite{Suh2004}.

\section{Static switching response}

\begin{figure}[!ht]
	\centering
	\begin{subfigure}[c]{0.49\linewidth}
		\centering
		\includegraphics[width=\linewidth]{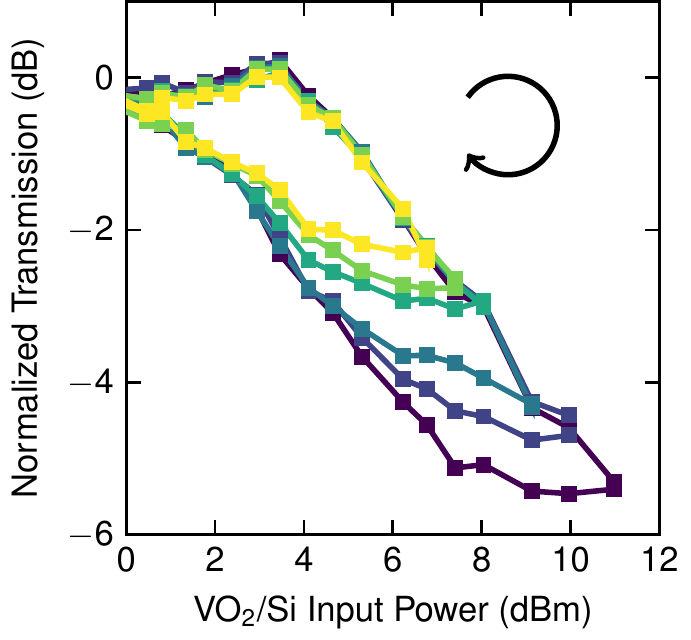}
		\caption{\centering}
		\label{fig:Cycles}
	\end{subfigure}	
	\hfill
	\begin{subfigure}[c]{0.49\linewidth}
		\centering
		\includegraphics[width=\linewidth]{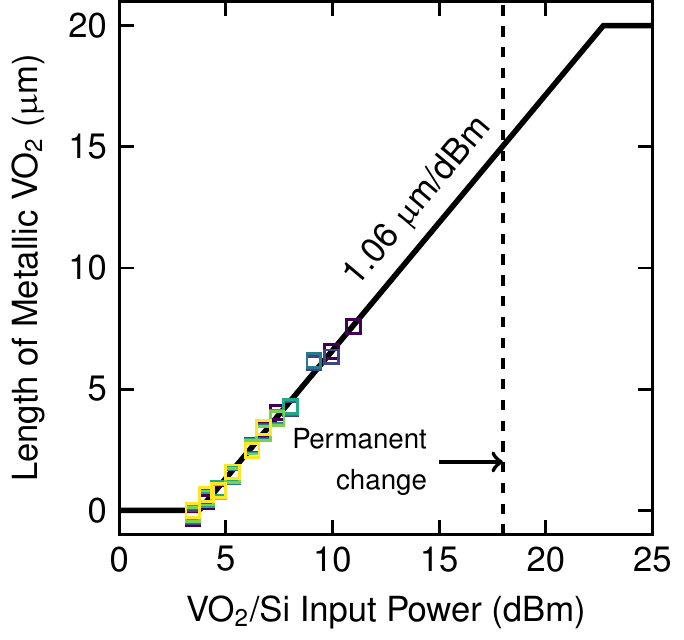}
		\caption{\centering}
		\label{fig:Damage}
	\end{subfigure}	
	\caption{(a) Normalized transmission as a function of the optical power. Different optical cycles were performed varying the maximum power. The arrow indicates the direction of the cycle. (b) Fragment of VO$_{2}$ in the metallic state as a function of the optical power from experimental results (square markers) and fitting (solid line). For both figures, optical power is given at the facet of the hybrid waveguide.}
\end{figure}

Prior to temporal measurements, the all-optical static response of the hybrid waveguide was investigated. The transmission of the waveguide was collected as a function of the input power. A tunable continuous-wave (CW) laser with an erbium-doped fiber-amplifier (EDFA) was used to generate high power optical signals to drive the VO$_{2}$ into the metallic state. The normalized transmission as a function of the input power at the facet of the hybrid waveguide is shown in Fig. \ref{fig:Cycles}. The power at the facet is calculated by subtracting the output power delivered by the EDFA to the grating estimated loss from the Si reference waveguide (see Fig. \ref{fig:Spectra}). Several optical cycles were performed varying the maximum power. On one hand, a hysteresis is observed for all the cycles, which confirms that optical loss is due to the metallic/insulating change of the VO$_{2}$. On the other hand, the loss is proportional to the input power. This indicates that only a portion of the VO$_{2}$ patch becomes metallic and it is attributed to the attenuation that experiences the optical signal due to the VO$_{2}$ absorption. 

Based on these measurements and the simulated optical loss, the VO$_{2}$ underwent the metallic state with a rate of 1.06 $\upmu$m/dBm for input powers higher than 3.5 dBm (Fig. \ref{fig:Damage}). To change the 20-$\upmu$m-long VO$_{2}$ on top of the waveguide, an optical power of 22.4 dBm would be necessary at the facet. It is important to take into account that in the static regime the IMT is triggered by photo-thermal excitation \cite{Ryckman2012}. Hence, the VO$_{2}$ could be damaged or permanently changed to another phase like V$_{2}$O$_{5}$ if this surpasses 600 ºC \cite{Liu2016}. In our hybrid waveguide, a permanent change was obtained when the input power was increased up to 18 dBm. A larger insertion loss was measured while a degradation in a fragment of the VO$_{2}$ layer was also observed by optical inspection. Furthermore, the response showed in Fig. \ref{fig:Cycles} could not be replicated using the same input powers.

\section{All-optical switching dynamics}
Based on the all-optical static measurements and constraints, the temporal dynamics of the hybrid waveguide were investigated. To this end, the temporal characteristics were obtained from the hybrid waveguide transmission using a pump-probe technique in the telecommunications wavelength region.

\begin{figure}[!th]
	\centering
	\includegraphics[width=\linewidth]{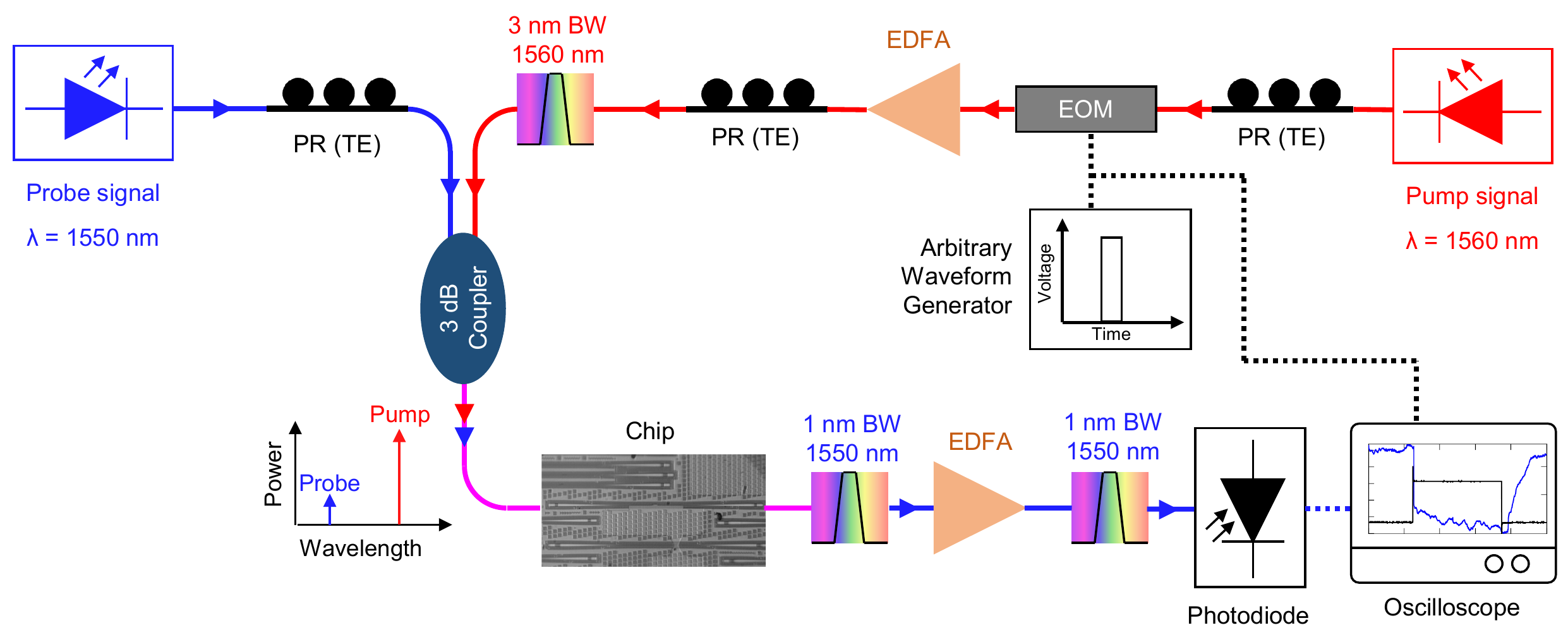}
	\caption{Schematic of the experimental set-up used for testing the all-optical temporal response of the hybrid VO$_{2}$/Si waveguide.}
	\label{fig:Set-up}
\end{figure}

The set-up used for this work is schemed in Fig. \ref{fig:Set-up}. On one hand, the probe signal was generated by using a CW laser at 1550 nm with an output power of 3 dBm and set to TE polarization with a 3-paddle manual polarization rotator (PR). On the other hand, the pump pulses were generated from a CW laser at 1560 nm with an output power of 5 dBm together with an external electro-optic modulator (EOM) and an arbitrary waveform generator. The output of the EOM was amplified with an EDFA. Then, the high-power pump signal was TE-polarized with another PR and filtered in order to remove the noise introduced by the EDFA. Before injecting the probe and pump signals to the chip, both were combined using a 3 dB directional coupler. The output of the chip was filtered to remove the pump signal, amplified with another EDFA to compensate chip optical loss, and filtered again to remove the EDFA noise. Finally, the temporal response was recorded by means of a high speed photodiode and an oscilloscope.

\begin{figure}[!ht]
	\centering
	\begin{subfigure}[c]{0.4\linewidth}
		\centering
		\includegraphics[width=\linewidth]{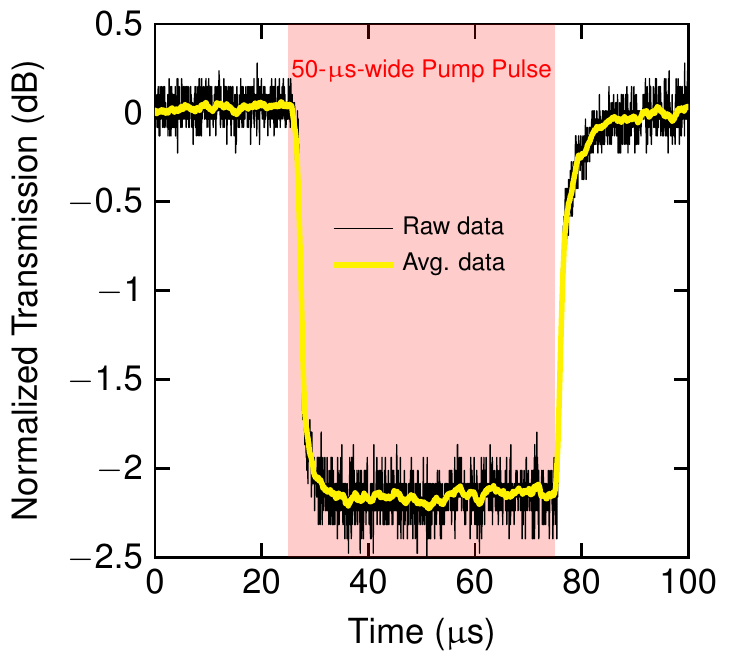}
		\caption{\centering}
		\label{fig:Pump-Probe-10kHz}
	\end{subfigure}	
	\hfill
	\begin{subfigure}[c]{0.4\linewidth}
		\centering
		\includegraphics[width=\linewidth]{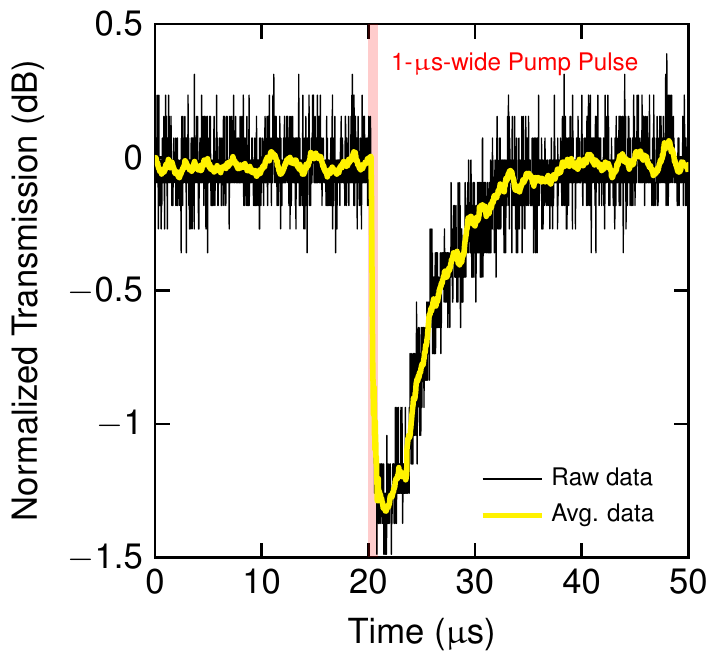}
		\caption{\centering}
		\label{fig:Pump-Probe-Pulse}
	\end{subfigure}	
	\\
	\begin{subfigure}[c]{0.4\linewidth}
		\centering
		\includegraphics[width=\linewidth]{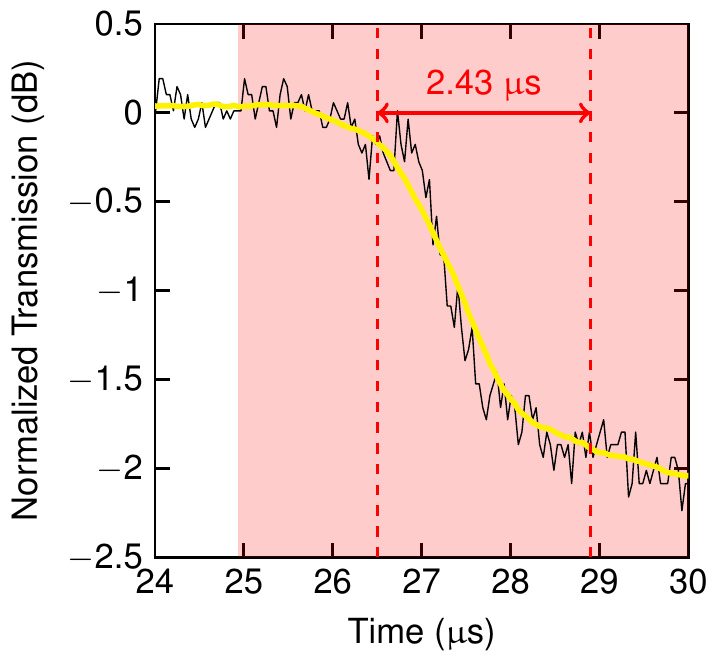}
		\caption{\centering}
		\label{fig:Pump-Probe-10kHz-IMT}
	\end{subfigure}	
	\hfill
	\begin{subfigure}[c]{0.4\linewidth}
		\centering
		\includegraphics[width=\linewidth]{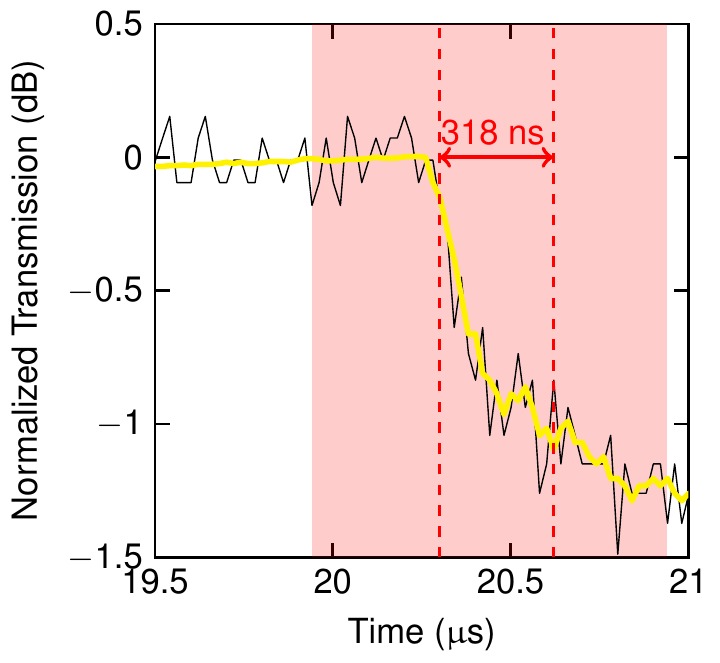}
		\caption{\centering}
		\label{fig:Pump-Probe-Pulse-IMT}
	\end{subfigure}	
	\\
	\begin{subfigure}[c]{0.4\linewidth}
		\centering
		\includegraphics[width=\linewidth]{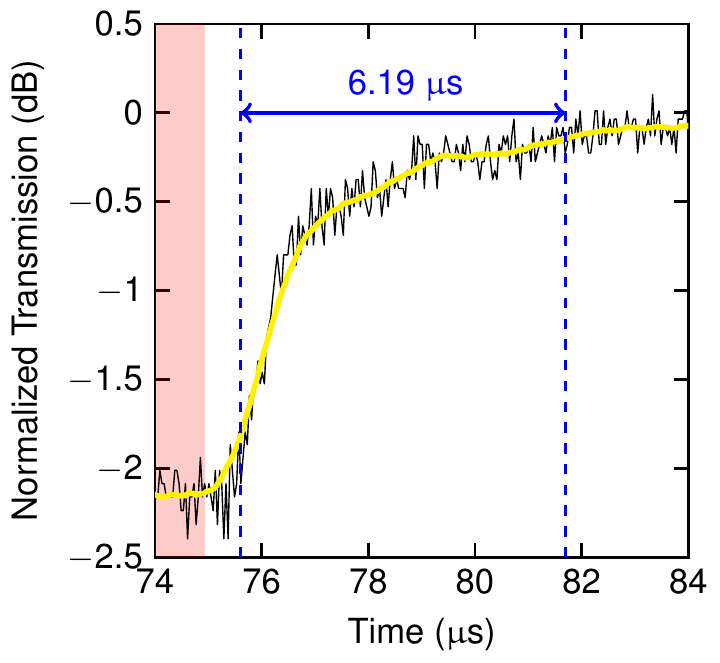}
		\caption{\centering}
		\label{fig:Pump-Probe-10kHz-MIT}
	\end{subfigure}	
	\hfill
	\begin{subfigure}[c]{0.4\linewidth}
		\centering
		\includegraphics[width=\linewidth]{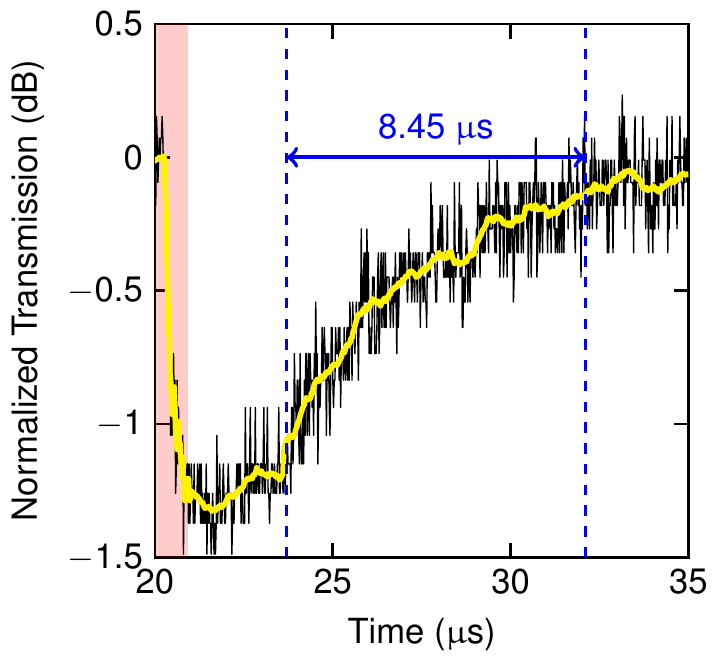}
		\caption{\centering}
		\label{fig:Pump-Probe-Pulse-MIT}
	\end{subfigure}	
	\caption{Normalized output of the probe signal as a function of the time upon: (a) 50-$\upmu$s-wide and 8 dBm rectangular pump pulse and (b) 1-$\upmu$s-wide and 12 dBm rectangular pump pulse. (c,e) and (d,f) are the zooms of fall and rise times for (a) and (b), respectively. Results are given at 1550 nm and 1560 nm for the probe and pump signals, respectively. The pump signal power is at the facet of the hybrid waveguide.}
\end{figure}

A 50-$\upmu$s-wide rectangular pulse with a repetition rate of 10 kHz was generated. The amplitude of the pump signal was set to obtain around 8 dBm at the facet of the hybrid waveguide. This lead to a energy consumption of 315 nJ. The normalized output is shown in Fig. \ref{fig:Pump-Probe-10kHz}. Upon the pump signal, the probe suffered a drop on its power in good agreement with the results obtained in the static measurements (see Fig. \ref{fig:Cycles}). The drop of the probe (OFF state) is associated with the high-loss metallic state of the VO$_{2}$, whereas the recovery (ON state) is linked to the low-loss insulating one. The switching times were obtained from the averaged data and following the 10\%-90\% rule. A value of 2.43 $\upmu$s (Fig. \ref{fig:Pump-Probe-10kHz-IMT}) and 6.19 $\upmu$s (Fig. \ref{fig:Pump-Probe-10kHz-MIT}) was obtained for the OFF and ON states, respectively. The difference between both values is attributed to the thermal increment required to complete the IMT and MIT of the VO$_{2}$ (see Fig. \ref{fig:Hysteresis}). The longer switch time of the ON state is due to the fact the MIT needs a wider switching window compared to the IMT (OFF state) to complete the phase transition.

Nonetheless, switching times are larger than the pico and femtosecond times reported in previous works using out of plane optical excitation \cite{Becker1996,Cavalleri2001,Cavalleri2004,Baum2007,Bionta2018,Haglund2019}. The differences in the timescales suggest a slower mechanism behind the VO$_{2}$, which could be attributed to: (i) the VO$_{2}$ layer is not uniformly illuminated due to in-plane approach; and (ii) the power of the pump signal is much lower. Notably, switching times are in the same order of magnitude to those reported using microheaters \cite{Sanchez2018}. Therefore, the switching mechanism is attributed to be of thermal origin. On one hand, the insulating to metal switching (OFF switching) is due to the fact of heat arising from the VO$_{2}$ optical absorption. On the other hand, the metal to insulating switching (ON switching) involves a thermal dissipation process.

The temporal response of the device is thus described by the heat conduction dynamics and conditioned by the surrounding materials as in thermo-optic phase shifters \cite{Parra2020}. Nevertheless, the nonlinear thermal temporal response can be taken in advantage to reduce the insulating to metal switching times while maintaining the extinction ratio. To this end, higher optical pump powers should be used together with a drastic shortening of the pulse width. The first, to increase the heating rate, while the second to avoid the VO$_{2}$ thermal damage threshold. To prove this point, the pulse width was 50 times decreased to 1 $\upmu$s whereas the pump power was only increased 2.5 times up to 12 dBm. These values were chosen with the aim of achieving a similar extinction ratio. The normalized temporal response is shown in Fig. \ref{fig:Pump-Probe-Pulse}. The switching time of insulating to metal was reduced one order of magnitude down to 318 ns (Fig. \ref{fig:Pump-Probe-Pulse-IMT}). This reduction is attributed to the higher heating rate, which was confirmed by observing the delay between the pump and the actual switch of the probe. For this latter case, the delay was reduced from around 1.5 $\upmu$s (Fig. \ref{fig:Pump-Probe-10kHz-IMT}) to 300 ns (Fig. \ref{fig:Pump-Probe-Pulse-IMT}). On the other hand, the metal to insulating switching time remained at a similar value (Fig. \ref{fig:Pump-Probe-Pulse-MIT}), since it only depends on the thermal dissipation rate of the structure \cite{Markov2015,Joushaghani2015}. Furthermore, the non-proportional shortening of the pulse width with the power increase leads to a reduction of the energy consumption, being for this case one order of magnitude lower and achieving 15.8 nJ. 

As a consequence, the switching times in all-optical hybrid VO$_{2}$/Si devices using an in-plane approach are governed by thermal conductive dynamics. Nonetheless, the switching time from insulating to metal state can be reduced by increasing the pump power while reducing the pump pulse, which also benefits of achieving lower energy consumption values. However, high optical powers may not be suitable in large-footprint photonic integrated circuits because of silicon two-photon absorption (TPA) and free carrier absorption (FCA)  \cite{Grillanda2015}. One solution could be to replace the silicon waveguide by a SiN one \cite{Moss2013}. On the other hand, the switching time from metal to insulating state could be reduced by using high diffusivity surrounding materials or making smaller the size of the VO$_{2}$ patch \cite{Ryckman2013}.

\section{Conclusions} 
In conclusion,  we have experimentally investigated the temporal dynamics of hybrid VO$_{2}$/Si waveguides upon an integrated and all-optical driven approach. On one hand, all-optical static results show a maximum of 5 dB of ER in a 7.5-$\upmu$m-long VO$_{2}$ fragment for TE polarization. This value is limited by the maximum temperature that can achieve the VO$_{2}$ without suffering a permanent and irreversible change. On the other hand, all-optical temporal results show that the temporal dynamics of this device are thermal, in which the heat arises from the light absorbed by the VO$_{2}$. These new understandings open the door for developing future all-optical VO$_{2}$ photonic integrated devices with nanosecond switching speed.\\

%\begin{acknowledgement}
\textbf{Acknowledgement.} The authors thank David Zurita for his help with the experimental set-up, and Maria Recaman for her inputs with VO$_{2}$. \\
%\end{acknowledgement}

%\begin{funding}
\textbf{Funding.} Ministerio de Economía y Competitividad (MINECO) (TEC2016-76849); Ministerio de Ciencia e Innovación (PID2019-111460GB-I00, FPU17/04224); Generalitat Valenciana (PROMETEO/2019/123).
%\end{funding}

\bibliographystyle{unsrt}
\bibliography{Bib}

\end{document}